\documentclass[
prd,
superscriptaddress,
10 pt,
twocolumn,
preprintnumbers,
notitlepage,
secnumarabic,
nobibnotes,
nofootinbib,
showpacs
]{revtex4-1}

\usepackage[T1]{fontenc}
\usepackage[pdftex]{color,graphicx}
\usepackage{mathtools}
\usepackage{amssymb}
\usepackage{bm}
\usepackage{dsfont}
\usepackage{mathrsfs}
\usepackage{calligra}
\usepackage{tabularx}
\usepackage[artemisia]{textgreek}
\usepackage{hyperref}
\usepackage{xcolor}
\usepackage{enumerate}
\usepackage{multirow}
\usepackage{url}
\usepackage{caption}
\usepackage{ulem}
\usepackage{perpage}
\MakePerPage{footnote}
\renewcommand{\.}{\!\;}		%	2 point space
\renewcommand{\@}{\!\:\!}	%	-2 point space

\DeclareMathAlphabet\mathbfcal{OMS}{cmsy}{b}{n}

\newcommand{\pos}[1]{\ensuremath{\@\langle#1\rangle}}
\begin{document}
\title{Lattice classical cosmology}

\author{Jakub Bilski}
\email{bilski@zjut.edu.cn}
\affiliation{Institute for Theoretical Physics and Cosmology, Zhejiang University of Technology, 310023 Hangzhou, China}

	%%%%%%%%%	%%%%%%%%%	%%%%%%%%%	%%%%%%%%%

	%%%%%%%%%	%%%%%%%%%	%%%%%%%%%	%%%%%%%%%
\begin{abstract}
\noindent
This article presents the lattice-smeared gravity phase space reduction defined by the cosmological gauge-fixing conditions. These conditions are specified to reduce the SU$(2)$ symmetry and the spatial diffeomorphism invariance of the loop quantum gravity's Fock space, known as the spin network. The internal symmetry is fixed to the Abelian case and the diffeomorphism invariance is simultaneously reduced to spatial translations. The unification of the results of the related gauge fixing conditions leads to the gauge generators correlation. Consequently, these conditions become solvable by constant variables; hence the reduced constraints become globally satisfied and vanish identically. By rigorously satisfying the reduced gauge symmetries, the resulting cosmological model is precisely the limit of the gravitational theory expressed in terms of holonomies and fluxes. Moreover, the obtained Hamiltonian constraint is finite (without any cut-off introduction) and as rigorous as an approximation of a Lie group by its representation. Furthermore, it has the form of the sum over elementary cuboidal cells. Finally, the simple structure of its homogeneities and anisotropies should allow to describe the quantum cosmological evolution of the Universe in terms of transition amplitudes, instead of using perturbative approximations.
\end{abstract}
	%%%%%%%%%	%%%%%%%%%	%%%%%%%%%	%%%%%%%%%

\maketitle

	%%%%%%%%%	%%%%%%%%%	%%%%%%%%%	%%%%%%%%% 

	%%%%%%%%%	%%%%%%%%%	%%%%%%%%%	%%%%%%%%%
\section{Motivation}\label{Sec_Motivation}

\noindent
The idea to formulate a cosmological theory on a lattice was inspired by the recent results in the development of an analogous approach to gravity \cite{Bilski:2020xfq,Bilski:2020poi,Bilski:2021ysc,Bilski:2021hrr}. Consequently, the model described in terms of holonomies of the Ashtekar-Barbero connections and fluxes of densitized dreibeins \cite{Ashtekar:1986yd,Barbero:1994ap} was constructed. The same variables have been used in the formulation of loop quantum gravity (LQG) \cite{Thiemann:1996aw,Thiemann:2007zz}. However, the theory presented in this article is not formulated by assuming a simplified form of the Ashtekar variables, which describe cosmological degrees of freedom. It is neither constructed by applying this form 
to the LQG's variables.

The model, called loop quantum cosmology \cite{Ashtekar:2003hd,Ashtekar:2009vc}, based on this latter construction, requires a UV cut-off. Moreover, this cosmological theory is not defined as a quantum model of the phase space-reduced classical lattice gravity, which after the standard canonical quantization \cite{Dirac:1925jy,Heisenberg:1929xj} leads to LQG.

The strategy presented in this article is more rigorous. It concerns the standard lattice gauge theory approach \cite{Bilski:2021hrr} and the phase space reduction of the lattice-smeared canonical variables \cite{Bilski:2019tji}, which are the classical variables before the canonical quantization in the LQG procedure \cite{Thiemann:1996aw,Thiemann:2007zz}. As a direct result, the system with a finite Hamiltonian (without a cut-off) is obtained. Moreover, the canonical quantization of this lattice-derived phase space-reduced scalar constraint and the so-obtained canonical fields directly corresponds to the quantum lattice gravity \cite{Bilski:2021hrr}. In particular, the variables of the lattice cosmological model presented in this article satisfy the equivalently reduced gauge symmetries of the LQG's Fock space \cite{Thiemann:1996aw,Thiemann:1996av,Thiemann:1997rv}, known as the spin network.

	%%%%%%%%%	%%%%%%%%%	%%%%%%%%%	%%%%%%%%% 

	%%%%%%%%%	%%%%%%%%%	%%%%%%%%%	%%%%%%%%%
\section{Lattice gravity}\label{Sec_Observable}

\noindent
The ADM formulation of gravitation \cite{Arnowitt:1960es} imposes a particular gauge on the time reparametrization anisotropy. The unique direction of the spatial evolution allows to construct a quantity identified as the momentum in the standard canonical relation to the gravitational field representative. Considering a Yang-Mills-like \cite{Yang:1954ek} formulation of gravitation, the field is represented by the $\mathfrak{su}(2)$-valued connection $A_a:=A_a^i\tau_i$. The related momentum is given by the dentitized dreiben $E^a:=E^a_i\tau_i$. The gauge generators are normalized as
\begin{align}
\label{Lie_real}
[\tau_j,\tau_k]=\epsilon^i_{\ jk}\tau_i\,.
\end{align}
This ensures the reality of the pair of canonical fields, known as the Ashtekar variables. Their Poisson algebra regarding the original ADM variables \cite{Arnowitt:1960es} $q_{ab}$ and $p^{ab}$ reads
\begin{align}
\label{Poisson_Ashtekar}
\big\{A^i_a(x),E_j^b(y)\big\}_{\!\!_{q\@,p}\!\!}=-\frac{\gamma\kappa}{2}\delta_a^b\delta^i_j\.\delta^3\@(x-y)\,,
\end{align}
where the normalization of the Einstein constant is $\kappa:=16\pi G/c^4$ and $\gamma$ denotes the Immirzi parameter. This allows to define the symplectic structure of the spatial manifold concerning the $A_a$ and $E^a$ fields. Consequently, the Poisson brackets become $\{X,Y\}:=-\frac{\gamma\kappa}{2}\{X,Y\}_{\@\@_{A\@,E}}$.

The gravitational theory formulation, equivalent to the Einstein-Hilbert model and expressed in the Ashtekar variables, is given by the Holst action \cite{Holst:1995pc}. The Legendre transform leads to the total Hamiltonian without explicit dynamical contribution. This Hamiltonian consists of three first-class constraints that generate the $\mathfrak{su}(2)$, spatial diffeomorphisms, and time diffeomorphism symmetries. They are called the Gauss, diffeomorphism (or vector), and Hamiltonian (or scalar) constraints. The last element is $H={\@\int\!d^3x\.N(x)\.\mathcal{H}(x)}$, where $N$ is the Lagrange multiplier called the lapse function. The solution of this constraint reveals the dynamics of the system. Therefore, one can first derive the Gauss and vector constraints. This leads to the physical subspace of the phase space and it is the constraints-solving order in the lattice regularization procedure in the LQG construction. In this case, one defines the $\mathfrak{su}(2)$- and diffeomorphisms-invariant Fock space \cite{Thiemann:1996aw,Thiemann:1996av,Thiemann:1997rv} founded on the lattice, on which the scalar constraint operator acts. This operator is formulated by regularizing the Hamiltonian constraint density
\begin{align}
\label{Holst_Hamiltonian}
\mathcal{H}:=
\frac{1}{\kappa}
|E|^{-\frac{1}{2}}\big(F^i_{ab}-(\gamma^2\@+1)\epsilon_{ilm}K^l_aK^m_b\big)\epsilon^{ijk}E_j^aE_k^b\,,
\end{align}
where $E$ is the densitized dreibein determinant, $F_{ab}$ is the curvature of $A_a$, and $K_a$ denotes the dreibein-contracted extrinsic curvature.

The regularization procedure depends on a particular choice of the graph on which the lattice is founded by embedding in the spatial manifold. In the established formulation of LQG \cite{Thiemann:1996aw,Thiemann:2007zz}, one assumes a tetrahedral tessellation of the spatial manifold. This choice, however, does not reflect the Cartesian symmetry of the differential form in the following transition from the continuous into a discrete framework,
\begin{align}
\label{integral_sum}
\int\!\!d^3x\.\text{f}(x)=\lim_{\bar{l}\to0}\sum_{R_{c(\@v\@)}\!\!\!}\text{f}\big(\@R_{c(\@v\@)}\@\big)\.\bar{l}^{\,3}\pos{v}\,.
\end{align}
Here, ${\bar{l}\.\pos{v}}:={\mathbb{L}_0\.\bar{\varepsilon}\.\pos{v}}$ is the elementary cell volume, where the isotropic regulator is given by the relation ${\bar{\varepsilon}^{\,3}\pos{v}}:=\epsilon^{(\@p\@)\@(\@q\@)\@(\@r\@)}{\varepsilon_{(\@p\@)}\pos{v}}\.{\varepsilon_{(\@q\@)}\pos{v}}\.{\varepsilon_{(\@r\@)}\pos{v}}$. The quantity $\mathbb{L}_0$ is an auxiliary length scale and ${\varepsilon_{(\@p\@)}\pos{v}}$ denotes the regularization parameter assigned to a particular link. The indices written in brackets are not summed. The summation in \eqref{integral_sum} goes over elementary cells $R_{c(\@v\@)}$, the edges of which, ${\partial R_{c(\@v\@)}}\in\Gamma$, are identified with the $\Gamma$ graph's edges, where ${c(v)}:=v\@+\@l_x\@/2\@+\@l_y\@/2\@+\@l_z\@/2$ denotes the $R$'s centroid.

The mentioned transition problem, indicated in \cite{Bilski:2020poi}, led to a different selection of the regularization framework. This problem (as well as several other constructional shortcomings of LQG) was resolved by choosing the lattice over the $\Gamma$ graph defined on the edges of the quadrilaterally hexahedral honeycomb \cite{Bilski:2020poi,Bilski:2021hrr}. Consequently, the starting point toward the formulation of the cosmological framework is the relativistic gravitational theory on the quadrilaterally hexahedral lattice \cite{Bilski:2021hrr,Zapata:1997da}.

In this model, likewise in the former theory in \cite{Thiemann:1996aw,Thiemann:2007zz}, the connection's degrees of freedom are smeared along the graph's edges and are represented by holonomies
\begin{align}
\label{holonomy}
h_{p}^{-1}\pos{v}[A]:=\mathcal{P}\exp\!\bigg(\!-\!\int_{\@l_p\pos{v}}\!\!\!\!\!\!\!\!ds\,\dot{\ell}^a\@(s)\.A_a\@\big(\ell(s)\big)\!\bigg).
\end{align}
The densitized dreibeins are expressed in terms of the fluxes through the surfaces orthogonal to their spatial directions. For the hexahedral cell's face $F^p\pos{v}$, oriented according to the right-hand rule-related multiplication $l_{q\@}\wedge l_r$, the flux is defined as
\begin{align}
\label{flux}
f\@\big(\@F^p\pos{v}\@\big)[E]
:=\epsilon^{pqr}\!\int_{\@l_q\pos{v}}\!\!\!\!\!\!\!\!dy\int_{\@l_r\pos{v}}\!\!\!\!\!\!\!\!dz\.\epsilon_{bcd}E^b\partial_yx^c\partial_zx^d\,.
\end{align}
The holonomies-and-fluxes-dressed graph is going to be named the lattice and its edges ${l_p\pos{v}}$ emanated from points $v$ (called nodes), are going to be named links.

The lattice smearing of the canonical variables adjusts the spatial directions $x_a$ into graph's edges $l_p$ (that become links) and results in
\begin{align}
\label{lattice_smearing}
\begin{split}
A_a(x)&\to\mathtt{A}(l_p)
:=A_p\@\big[h_{(\@p\@)}\big]\,,
\\
E^a(x)&\to\mathtt{E}(F^p)
:=E^p\@\big[f^{(\@p\@)}\big]\,.
\end{split}
\end{align}
This map is constructed by the expansion of definitions \eqref{holonomy}, \eqref{flux} and inversion of the obtained results. However, to preserve the symmetry of cells $R_{c(\@v\@)}$, the point-located fields $A_a(x)$ and $E^a(x)$ must be equally distributed around the $c\.\pos{v}$ centroids of quadrilateral hexahedra. The simplest method is to integrate their distribution densities along the spatial directions that are not present in \eqref{holonomy} and \eqref{flux},
\begin{align}
\label{cell_symmetrizing}
\begin{split}
\bar{\mathtt{A}}_{p\@}\big(\@R_{c(\@v\@)}\@\big)
:=&\int_{\!F^{(\@p\@)}\pos{v}}\!\!\!\!\!\!\!\!\!\!\!\!\mathcal{A}_p\@\big[h_{(\@p\@)}\big]
=\frac{\varepsilon_{(\@p\@)}\pos{v}}{\mathbb{L}_0^2\.\bar{\varepsilon}^{\,3}\pos{v}\!}
\int_{\!F^{(\@p\@)}\pos{v}}\!\!\!\!\!\!\!\!\!\!\!\!A_p\@\big[h_{(\@p\@)}\big]\,,
\\
\bar{\mathtt{E}}^{p\@}\big(\@R_{c(\@v\@)}\@\big)
:=&\int_{\@l_{(\@p\@)}\pos{v}}\!\!\!\!\!\!\!\!\!\!\mathcal{E}^p\@\big[f^{(\@p\@)}\big]
=\frac{1}{\mathbb{L}_0\.\varepsilon_{(\@p\@)}\pos{v}\!}
\int_{\@l_{(\@p\@)}\pos{v}}\!\!\!\!\!\!\!\!\!\!E^p\@\big[f^{(\@p\@)}\big]\,.
\end{split}
\end{align}
These expressions formally assign the canonical variables to the lattice in the manner both symmetric with its structure and consistent with the transition in \eqref{integral_sum}.

	%%%%%%%%%	%%%%%%%%%	%%%%%%%%%	%%%%%%%%% 

	%%%%%%%%%	%%%%%%%%%	%%%%%%%%%	%%%%%%%%%
\section{Ashtekar-Barbero connection as a candidate for an observable}\label{Sec_Observable}

\noindent
Considering a quantum theory formulated in terms of the SU$(2)$ holonomies of the $\mathfrak{su}(2)$ Ashtekar connections, these fields reflect the group-representation duality \cite{Bilski:2020xfq}. Moreover, the Hamiltonian structure in \eqref{Holst_Hamiltonian} is similar to its Yang-Mills analog \cite{Yang:1954ek}. Therefore, it is natural to apply the same quantization strategy regarding the pair $A_p$ and $h_p$ as the one concerning gauge fields in the Standard 
Model of particle physics.

By following the analysis in \cite{Bilski:2020xfq}, one chooses the probability distribution ${\mathtt{A}(\@l_{(\@p\@)\@}\pos{v}\!):=\mathtt{A}_{(\@p\@)}\pos{v}\@:=\theta_{(\@p\@)}\pos{v}\!/\mathbb{L}_0\.\varepsilon_{(\@p\@)}\pos{v}}$ of the gravitational connection, introduced in \eqref{lattice_smearing}. Here, this quantity has been redefined in terms of the representation
\begin{align}
\label{abstract_representation}
\theta_{(\@p\@)}\pos{v}:=\theta_{(\@p\@)}^i\pos{v}\.\tau^i
:=\!\int_0^{\mathbb{L}_0\.\varepsilon_{(\@p\@)}\@\pos{v}}\!\!\!\!\!\!\!\!ds\.\dot{\ell}^a\@(s)\.A_a\@\big(\ell(s)\big)
\end{align}
of the gauge group determined by the holonomy in \eqref{holonomy}. The power series representation of the gauge transformation reads
\begin{align}
\label{connection_distribution}
\mathtt{A}_{(\@p\@)}\pos{v}
=\frac{1}{\mathbb{L}_0\.\varepsilon_{(\@p\@)}\pos{v}\!}
\sum_{n=0}^{\infty}\frac{1}{n!}\frac{\mathrm{d}^{\!\.n}\theta_{(\@p\@)}\pos{v}\!}{\mathrm{d}\varepsilon_{(\@p\@)}^n\@\pos{v}\!}
\bigg|_{\@\varepsilon_{(\@p\@)}\@\pos{v}\!\.=\,\@0}\varepsilon_{(\@p\@)}^{n\mathstrut}\pos{v}\,,
\end{align}
where the constraints on the value of the regulator have to be specified as follows,
\begin{align}
\label{regulator}
\forall_p\forall_v\ 0\leq\varepsilon_{(\@p\@)}\pos{v}:=\frac{l_{(\@p\@)}\pos{v}\@}{\mathbb{L}_0}\ll1\,.
\end{align}
Then, the expansion of the group element around the identity up to the quadratic-order terms reads
\begin{align}
\begin{split}
\label{holonomy_exponential_expansion}
h_{(\@p\@)}^{\mp1}\pos{v}\@[\,\@\theta\,\@]=&\;\mathds{1}
+\mathbb{L}_0\.\varepsilon_{(\@p\@)}\pos{v}\mathtt{A}_{(\@p\@)}\pos{v}
\\
&+\frac{1}{2}\mathbb{L}_0^2\.\varepsilon_{(\@p\@)}^2\pos{v}\mathtt{A}^2_{(\@p\@)}\pos{v}
+\mathcal{O}\big(\varepsilon_{(\@p\@)}^3\pos{v}\big)\,.
\end{split}
\end{align}
Here, the equality ${\mathcal{O}\big(\theta_{(\@p\@)}^3\pos{v}\big)}={\mathcal{O}\big(\varepsilon_{(\@p\@)}^3\pos{v}\big)}$, demonstrated in \cite{Bilski:2020xfq,Bilski:2021hrr}, has been implemented. It is also worth noting that the term ${[\theta_{(\@p\@)}\pos{v}\@,\theta_{(\@p\@)}\pos{v}\@]}$, proportional to ${\varepsilon_{(\@p\@)}^2\pos{v}}$, has vanished due to the antisymmetry of the Lie brackets.

Finally, one needs to specify the smearing of the continuous and point-located connection ${A_{(p)}\@(x)}$ into a discrete and link-distributed probability ${\mathtt{A}_{(\@p\@)}\pos{v}}$. This issue regarding a general model of lattice gravity is elaborated in \cite{Bilski:2021hrr}.

	%%%%%%%%%	%%%%%%%%%	%%%%%%%%%	%%%%%%%%% 

	%%%%%%%%%	%%%%%%%%%	%%%%%%%%%	%%%%%%%%%
\section{Phase space reduction}\label{Sec_Reduction}

\noindent
The internal space of the Ashtekar variables is Euclidean. The same geometry is often considered in cosmology. The most general Euclidean geometry model, the homogeneous and isotropic limit of which has the Friedmann-Lema\^i{}tre-Robertson-Walker metric, is the Bianchi I cosmology. It can be constructed by imposing constraints on the form of the lattice gravity variables. Its simplest realization is using only the diagonal $A_a^i$ and $E^a_i$ coefficients \cite{Ashtekar:2009vc}. This assumption can be formally derived as a solution of the phase space reduction of the lattice-assigned variables in \eqref{lattice_smearing}, \textit{cf.} \cite{Bilski:2019tji}. In this construction, the gauge-fixing conditions are imposed globally. The $\mathfrak{su}(2)$ invariance is broken into the Abelian case, and the spatial diffeomorphism invariance is analogously restricted into linear translations. This latter reduction leads to the lattice composed of linear links --- all of them are either orthogonal or parallel one to another. Thus, the piecewise linear structure of the lattice gravity in \cite{Bilski:2021ysc,Bilski:2021hrr,Zapata:1997da} becomes constrained to the rectangularly hexahedral (cuboidal) form. It is worth noting that the piecewise analytical structure of the lattice in LQG \cite{Thiemann:1996aw,Thiemann:2007zz} would be constrained to the same form.

The advantages of this formal phase space reduction \cite{Bilski:2019tji} are revealed in quantization. Dirac brackets, restricted by the gauge conditions to the reduced phase space variables, turn into Poisson brackets. This ensures that the reduced quantum theory remains the gauge-fixed variant of the general model and no information is lost. Concerning this analysis, the semiclassical limit of the so-constructed lattice cosmology equals the reduced limit of the related quantum gravity. It is worth emphasizing that to preserve this equality, the global reduction must restrict also the states supporting the quantum evolution and any perturbative analysis.

The reduced phase space fields are indeed diagonal \cite{Bilski:2019tji}. It is useful to express these diagonal coefficients of the Ashtekar variables in the following normalization \cite{Ashtekar:2003hd}:
\begin{align}
\label{reduction}
\begin{split}
\mathtt{A}(l_p)
&\to
c_i\@\big(l_{(\@i\@)}\big)\@[h]\tau_i
:=\mathbb{L}_0\varepsilon_{(\@i\@)}\pos{v}
A_p^i\@\big[h_{(\@p\@)}\big]{}^{\scriptscriptstyle0\@\@}e_{\@(\@i\@)}^p\tau_i\,,
\\
\mathtt{E}(F^p)
&\to
p^i\@\big(F^{(\@i\@)}\big)\@[f]\tau_i
:=\frac{\mathbb{L}_0^2\.\bar{\varepsilon}^{\,3}\pos{v}\!}{\varepsilon_{(\@i\@)}\pos{v}}
E^p_i\@\big[f^{(\@p\@)}\big]\frac{{}^{\scriptscriptstyle0\@\@}e^{\@(\@i\@)}_p}{\!\sqrt{{}^{\scriptscriptstyle0\@\@}q}}\tau_i\,.
\end{split}
\end{align}
The diagonal constant unit matrices ${}^{\scriptscriptstyle0\@\@}e_{i}^p$ and ${}^{\scriptscriptstyle0\@\@}e^{i}_p$ are the projector from the internal space basis into lattice directions and its reciprocal, respectively. The square root of the diagonal constant metric tensor determinant has the standard frame fields representation,
$\sqrt{{}^{\scriptscriptstyle0\@\@}q}:=\epsilon_{ijk}\epsilon^{pqr}\.{}^{\scriptscriptstyle0\@\@}e^i_p{}^{\scriptscriptstyle0\@\@}e^j_q{}^{\scriptscriptstyle0\@\@}e^k_r/3!\.$.

The solutions of the gauge fixing conditions determine the linear maps in \eqref{reduction} between the basis elements of internal and external spaces. These maps are constant along each indistinguishable direction \cite{Bilski:2019tji}. Thus, they result in the identical vanishing of both reduced constraints. The Abelian Gauss symmetry and the spatial translations invariance are no longer local gauge symmetries but become global features of constant variables defined along Euclidean directions.

Finally, by expanding the properly selected combinations (\textit{cf.} \cite{Bilski:2020poi,Bilski:2021hrr}) of the lattice quantities in \eqref{holonomy} and \eqref{flux}, restricted to the reduced variables $c_i(l_{(\@i\@)})[h]$ and $p^i(F^{(\@i\@)})[f]$, respectively, one obtains
\begin{align}
\label{reduced}
\begin{split}
h_{i}^{\mathstrut}\pos{v}[c]\@-\@h_{i}^{-1}\pos{v}[c]=&\;2c_{i}\@\big(l^{(\@i\@)}\big)\tau_{(\@i\@)}\,,
\\
f\@\big(\@F^{i}\pos{v}\@\big)[p]
=&\;p^{i}\@\big(F^{(\@i\@)}\big)\tau_{(\@i\@)}\,.
\end{split}
\end{align}
This is the lattice representation of the canonical pair of the reduced variables. In the upper equation, the higher-order terms from the exponential map expansion in \eqref{holonomy_exponential_expansion} were omitted. In the lower formula in \eqref{reduced}, the probability distribution was integrated with the trapezoidal rule for the constant densitized dreibein along a linear path.

It is worth to recall that the Poisson brackets of the  continuous analogs of the reduced variables (corresponding to \eqref{Poisson_Ashtekar}) are $\{c_i,p^j\}=-\frac{\gamma\kappa}{2}\delta_i^j$ \cite{Ashtekar:2003hd}. These, however, are not the variables related to the lattice phase space reduction. The Poisson algebra of the canonical pair in \eqref{reduction} is
\begin{align}
\label{Poisson_reduced}
\begin{split}
&\:\big\{c_i\@(l\pos{v})\@[h],p^j\@(F\pos{v'})\@[f]\big\}
\\
=&\ \text{tr}\@\bigg(\!\tau_{(\@i\@)}\Big\{
f^{(\@j\@)}\@\big(\@F^j\pos{v'}\@\big)[p],h_{i}^{\mathstrut}\pos{v}[c]\@-\@h_{i}^{-1}\pos{v}[c]
\Big\}\!\bigg)
\\
=&\ \text{tr}\@\bigg(\!\Big\{
f\@\big(\@F^{(\@i\@)}\pos{v'}\@\big)[p],h_{(\@i\@)}^{\mathstrut}\pos{v}[c]\@-\@h_{(\@i\@)}^{-1}\pos{v}[c]
\Big\}\!\bigg)\delta_i^j
\\
=&\;\gamma\kappa\,\text{tr}\@\Big(\@\tau_{(\@i\@)}
\big(h_{(\@i\@)}^{\mathstrut}\pos{v}[c]\@-\@h_{(\@i\@)}^{-1}\pos{v}[c]\big)\delta_{l\pos{v}l\pos{v'}}\@\Big)\delta_i^j
\\
=&\,-\gamma\kappa\.c_{(\@i\@)}\@\big(l^{(\@i\@)}\big)\tau_{(\@i\@)}\delta_{l\pos{v}l\pos{v'}}\delta_i^j\,.
\end{split}
\end{align}
As expected, this result is the Abelian limit of the equivalent Poisson brackets of the gravitational variables \cite{Bilski:2021ysc,Bilski:2021hrr}.

	%%%%%%%%%	%%%%%%%%%	%%%%%%%%%	%%%%%%%%% 

	%%%%%%%%%	%%%%%%%%%	%%%%%%%%%	%%%%%%%%%
\section{Lattice cosmology}\label{Sec_Lattice}

\noindent
The cosmological fields $c_i\@\big(l_{(\@i\@)}\big)\@[h]$ and $p^i(F^{(\@i\@)})\@[f]$ are link- and face-defined objects, respectively. They do not reflect the cuboidal lattice symmetry from the point-related perspective of the continues domain of the Ashtekar variables. This problem can be easily resolved in the same way as in the general theory  \cite{Bilski:2021ysc} --- see formulas \eqref{cell_symmetrizing}. Consequently, we define the cell-symmetrized probability distributions of the reduced fields
\begin{align}
\label{cell_symmetrization}
\begin{split}
\bar{c}_i\@\big(\@R_{c(\@v\@)}\@\big)\@[h]
:=&\;\mathbb{L}_0\varepsilon_{(\@i\@)}\pos{v}\!
\!\int_{_{\scriptstyle\!\!F^{(\@p\@)}\pos{v}}}\!\!\!\!\!\!\!\!\!\!\!\!\mathcal{A}_p^i\@\big[h_{(\@p\@)}\big]
{}^{\scriptscriptstyle0\@\@}e_{\@(\@i\@)}^p\,,
\\
\bar{p}_i\@\big(\@R_{c(\@v\@)}\@\big)\@[f]
:=&\;\frac{\mathbb{L}_0^2\.\bar{\varepsilon}^{\,3}\pos{v}\!}{\varepsilon_{(\@i\@)}\pos{v}}
\!\int_{_{\scriptstyle\!l_{(\@p\@)}\pos{v}}}\!\!\!\!\!\!\!\!\!\!\mathcal{E}^p_i\@\big[f^{(\@p\@)}\big]
\frac{{}^{\scriptscriptstyle0\@\@}e^{\@(\@i\@)}_p}{\!\sqrt{{}^{\scriptscriptstyle0\@\@}q}}\,.
\end{split}
\end{align}
Analogously to the derivation of the lower equation in \eqref{reduced}, the densities are integrated with the trapezoidal rule, resulting in
\begin{align}
\label{cell_symmetrized}
\begin{split}
\bar{c}_i\@\big(\@R_{c(\@v\@)}\@\big)\@[h]
=&\;\frac{1}{4}\!\sum_{v\in F^{(\@i\@)}\pos{v}}\!\!\!\!\!
\text{tr}\@\Big(\tau^{(\@i\@)}\big(h_{i}^{-1}\pos{v}[c]\@-\@h_{i}\pos{v}[c]\big)\Big)\,,
\\
\bar{p}_i\@\big(\@R_{c(\@v\@)}\@\big)\@[f]
=&\;\frac{1}{2}\!\sum_{v\in l_{(\@i\@)}\pos{v}}\!\!\!\!
f^{(\@i\@)}\@\big(\@F^i\pos{v}\@\big)[p]\,.
\end{split}
\end{align}
These objects are the cell-related distributions of the phase space-reduced Ashtekar variables. They are the fundamental expressions of canonical fields in the lattice cosmology.

Eventually, the Hamiltonian constraint density corresponding to a single cell of $\Gamma$, formulated in terms of the fields derived in \eqref{cell_symmetrized}, equals
\begin{align}
\label{cell_Hamiltonian}
\begin{split}
\mathcal{H}\big(\@R_{c(\@v\@)}\@\big)
=&\,-\frac{1}{\gamma^2\kappa}
\frac{\bar{p}^i\@(R)\.\bar{p}^j\@(R)}{(|\bar{p}^1\@(R)\.\bar{p}^2\@(R)\.\bar{p}^3\@(R)|)^{\@\frac{1}{2}}\!}
\\
&\,\times\@\big(\bar{c}_i\@(R)\.\bar{c}_j\@(R)\@-\@\bar{c}_j\@(R)\.\bar{c}_i\@(R)\big)\,.
\end{split}
\end{align}
The Gauss and spatial diffeomorphism constraints, written in terms of constant variables, vanish identically. This confirms the correctness of the imposition of the gauge-fixing conditions \cite{Bilski:2019tji}.

Then, the total lattice scalar constraint is given by the elementary cell summation
\begin{align}
\label{total_Hamiltonian}
H(\Gamma)=\lim_{\bar{l}\to0}\sum_{R_{c(\@v\@)}\!\!\!}
N\big(\@R_{c(\@v\@)}\@\big)\mathcal{H}\big(\@R_{c(\@v\@)}\@\big)\,.
\end{align}
It is worth noting that ${\mathcal{H}\big(\@R_{c(\@v\@)}\@\big)}\sim{\mathbb{L}_0^3\.\bar{\varepsilon}^{\,3}\pos{v}}$, thus this result corresponds directly to the transition from the continuous into a discrete framework in \eqref{integral_sum}. The Hamiltonian constraint is finite, hence it does not require any additional regularization or a cut-off. It expresses the entire kinematics of the gravitational degrees of freedom in the lattice cosmology. The isotropic inhomogeneities are represented by different lapse function elements $N\big(R_{c(\@v\@)}\big)$. These quantities determine the density of links intersections concerning different positions along the coordinate time direction. The anisotropies are much more strictly constrained by the cuboidal structure of the $\Gamma$ graph, \textit{i.e.} the face-adjacent cells are restricted to have parallel nodes-adjacent links.

	%%%%%%%%%%%%	%%%%%%%%%%%%	%%%%%%%%%%%%

	%%%%%%%%%%%%	%%%%%%%%%%%%	%%%%%%%%%%%%
\section*{Acknowledgements}

\noindent
This research was partially supported by the National Natural Science Foundation of China grants Nos. 11675145 and 11975203.
The author thanks Anzhong Wang for constructive comments that substantially helped improving the article.

	%%%%%%%%%	%%%%%%%%%	%%%%%%%%%	%%%%%%%%%

	%%%%%%%%%	%%%%%%%%%	%%%%%%%%%	%%%%%%%%%

%\newpage

\end{document}